\documentclass[reprint,prd,aps,superscriptaddress,showpacs,showkeys,nofootinbib,notitlepage ]{revtex4-1}
\usepackage{amsmath}
\usepackage{graphicx}
\usepackage{longtable}

\usepackage{color}

\newcommand{\be}{\begin{equation}}
\newcommand{\ee}{\end{equation}}
\newcommand{\bea}{\begin{eqnarray}}
\newcommand{\eea}{\end{eqnarray}}
\newcommand{\etal}{et al.}

\begin{document}

\bibliographystyle{apsrev}

\title{Probing the integrated Sachs-Wolfe effect using embedded lens models}

\author{B. Chen}
\email{bchen3@fsu.edu}

\affiliation{Research Computing Center, Department of Scientific Computing,
Florida State University, Tallahassee, FL 32306, USA, bchen3@fsu.edu}

\author{R. Kantowski}
\email{kantowski@ou.edu}
\affiliation{Homer L.~Dodge Department~of  Physics and Astronomy, University of
Oklahoma, 440 West Brooks,  Norman, OK 73019, USA}

\date{\today}

\begin{abstract}
The photometry  profile of the integrated Sachs-Wolfe (ISW) effect  recently obtained by the \emph{Planck} consortium  by stacking patches of Cosmic Microwave Background (CMB) sky maps around a large number of cosmic voids, contains  a cold ring at about half the void's effective radius surrounded by a hot ring near the void's boundary.
The source of the temperature structure is assumed to be the ISW effect but the exact cause of the ringed structure is not currently well understood, particularly the outer hot ring.
Numerical simulations have suggested that hot/cold ring structures can be produced by  motions associated with nonlinear growths of cosmic structures whose gravitational potentials produce the ISW effect.
We have recently developed the embedded lens theory and the Fermat potential formalism which can be used to model the ISW effect caused by intervening individual lens inhomogeneities evolving arbitrarily.
This theory only requires knowledge of the void's projected mass profile as a function of the passing CMB photons' impact radius and the rate of change of that mass distribution at passage.
We present two simple embedded void lens models with evolving mass densities and investigate the ISW effect caused by these lenses.
Both  models posses expanding mass shells which produce hot rings around central cold regions, consistent with the recent observations.
By adding a small over-density at the void's center we can produce the slight positive temperature excess hinted at in \emph{Planck}'s photometric results.
We conclude that the embedded lens theory and the Fermat potential formalism is well suited for modeling the ISW effect.
\end{abstract}



\maketitle

\section{Introduction}

The {\it secondary} anisotropy of the Cosmic Microwave Background (CMB) caused by the gravitational potentials of inhomogeneities along the line of sight to the CMB photon's last scattering at redshift $z\sim1100$ is called the integrated Sachs-Wolfe (ISW) effect \cite{Sachs67}.
The ISW fluctuations caused by nonlinear growths of the density perturbations at low redshifts were investigated in \cite{Rees68} using Swiss cheese models \cite{Einstein45,Schucking54,Kantowski69} and their presence is referred to as the Rees-Sciama (RS) effect.
Following the pioneering work of \cite{Sachs67,Rees68} the ISW/RS effect has been studied extensively using analytical and numerical techniques such as exact general relativity (GR) modeling, approximate perturbation modeling, and N-body simulations \cite{Dyer76, Birkinshaw83, Gurvits86, Thompson87, Martinez90, Seljak96, Cooray02, Inoue06, Schafer06, Smith09, Cai10, Cai14, Hernandez10, Nadathur12, Ilic13, Merkel13, Finelli14, Szapudi14}.
For example \cite{Dyer76} studied the RS effect  for galaxy clusters using a simple Swiss cheese model (an expanding homogeneous  dust sphere) whereas \cite{Cooray02} related some nonlinear growth effects to the divergence of the large scale structure's momentum field (see also \cite{Schafer06, Merkel13}).
This second order nonlinear ISW contribution is similar to that caused by the transverse motion of a gravitational lens, i.e., the Birkinshaw-Gull effect \cite{Birkinshaw83, Gurvits86}.
The numerical modeling is usually based on the merging halo model for large scale structure clustering \cite{Scherrer91, Sheth97, Ma00}.
Understanding the ISW/RS effect can be considered critical to modern cosmology.
For example,  the ISW effect probes the evolutionary history of cosmic structures and the dynamics of dark energy \cite{Crittenden96}.
Detecting this effect will directly probe the negative pressure nature of the dark energy and complement geometrical probes such as SNe Ia as standard candles \cite{Riess98, Perlmutter99} and baryonic acoustic oscillations as standard rulers \cite{Eisenstein05}.
Accurately measuring the ISW/RS effect is also important for quantifying the non-Gaussian signatures in the primordial density fluctuations \cite{Planck14a,Planck14b,Planck15}.
Because of the equivalence principle the ISW effect is manifested as a frequency independent temperature shift in the CMB spectrum. It has been detected as correlations between the CMB temperature sky maps and tracers of large scale structures \citep{Boughn04, Ho08}.
It has also been detected by stacking patches of CMB sky maps around known large scale structures (galaxy clusters and cosmic voids). 
The ISW effect has been detected by several groups using this method \cite{Granett08,Planck14b}; however, the signal seems to be much  larger than predicted by both linear growth theory and numerical simulations \cite{Maturi07, Hernandez10, Nadathur12, Ilic13, Cai10, Cai14, Hotchkiss15}.
One interesting result is the strange shape of the CMB temperature profile of stacked cosmic voids, in particular, the expected cold central region is surrounded by an unexpected hot ring in the outer part of the profile \citep{Ilic13, Planck14a, Cai14}.
This structure is hard to explain within the framework of the linear ISW effect in a standard $\Lambda$CDM cosmology (see Fig. 9 of \cite{Planck14a}).
The stacked photometric profiles also show a hint of a small positive excess (at about $2\,\sigma$) when small filter radii ($\lesssim $ 20\% of the effective void radius) are used.
The \citet{Planck14a} suggested small over-densities near the centers of the stacked cosmic voids (intrinsic to the void-finding algorithm ZOBOV used for the void catalog \cite{Neyrinck08,Sutter12}) as the possible cause of this intriguing feature.

We have developed the embedded lens theory in recent papers \cite{Kantowski10, Kantowski12, Kantowski13, Chen10, Chen11, Chen13a, Chen13b, Kantowski14} using the Swiss cheese models.
We have introduced the concept of the {\it Fermat} potential (equivalent to the sum of the geometrical and potential part of the gravitational lensing time delay) for embedded lenses and have shown that the lowest order embedded lens theory can be obtained by applying a variational principle to the Fermat potential \cite{Kantowski13}.
By lowest order we mean small angle lensing caused by Newtonian perturbations of the background cosmology.
The source of these Newtonian perturbations are density variations, possible large in amplitude, but small in physical dimension compared to the lensing distances involved and small compared to the radius of the universe. They are also assumed to be slowly varying.
These constraints on the perturbations  exclude the necessity of including post-Newtonian corrections \cite{Will72}
in the gravity theory and result in an instantaneous Newtonian potential which satisfactorly describes transiting photon orbits.
These restrictions are precisely the same as those imposed in conventional linear lensing theory.
Our Fermat potential approach to lensing uses the projected lens mass density directly without ever computing the Newtonian potential.

By using the Fermat potential to formulate gravitational lensing the ISW effect produced by individual inhomogeneities can be obtained from a derivative with respect to the lens' redshift $z_d$ of the potential part of the lensing time delay $T_p$ \cite{Chen13a},
\be\label{dT}
\frac{\Delta {\cal T}}{\cal T} = H_d\frac{\partial T_p}{\partial z_d}
= \frac{H_dT_p}{1+z_d} + 2(1+z_d) \frac{r_{\rm s}H_d}{c} \int_{x}^{1}{\frac{dx'}{x'}\frac{\partial f(x',z_d)}{\partial z_d}},
\ee where $H_d=H(z_d)$ is the Hubble parameter at the lens redshift $z_d,$ and the potential part of the
time delay is defined by
\be\label{Tp}
cT_p(\theta_I,z_d) = 2(1+z_d)r_{\rm s}\int_{x}^{1}{\frac{f(x',z_d)-f_{\rm RW}(x')}{x'}dx'}.
\ee
In the above $\theta_I$ is the lensing image angle, $x=\theta_I/\theta_M$ is the normalized image angle, $\theta_M$ the angular radius of the comoving lens boundary, $r_{\rm s}$ is the Schwarzschild radius of the lensing inhomogeneity, and $f(x)\equiv M(x)/M$ is the (projected) fraction of the lens mass contained within the impact disk of radius $\theta_I,$ and $f_{\rm RW}(x) = 1-(1-x^2)^{3/2}$ is the corresponding quantity for the homogeneous Friedmann-Lema\^\i tre-Robertson-Walker (FLRW) sphere removed to form the embedded lens \citep{Chen13a}.
Embedded lensing differs from conventional lensing in that it accounts for the absence of the gravitational attraction of the removed Swiss cheese void'��s mass. 
The $f_{\rm RW} (x)$ term in Eq.\,(\ref{Tp}) is a consequence of embedding on time delays when linear lensing is adequate.
Embedding can produce percent size corrections to strong lensing time delays and image amplifications, and can produce similar size changes in weak lensing shears. It also naturally predicts the repulsive lensing produced  by cosmic voids \citep{Chen13b}.
Because the embedded lens mass is a contributor to the mean cosmic density its effect on passing photons vanishes at impacts $x\ge 1$.
For those photons that transit the lens the total CMB temperature fluctuation $\Delta{\cal T}$ can be split in two parts, the time-delay contribution $\Delta{\cal T}_T$ and the evolutionary contribution $\Delta {\cal T}_{\cal E},$ i.e., the two terms in Eq.~(\ref{dT}).
We have estimated the ISW signal caused by individual cosmic voids and clusters in \cite{Chen13a} considering only the time-delay contribution $\Delta {\cal T}_T$.
This may have over or under-estimated the ISW signal depending on the sign of the evolutionary contribution, see Eq.\,(10) of \cite{Chen13a}.
For example, in the case of linearly growing density perturbations, the evolutionary contribution cancels a significant part of the time-delay contribution.
Consequently, considering only the time-delay contribution $\Delta {\cal T}_{T}$ over-estimates the total effect.
On the other hand, for extremely nonlinear structures (e.g., for a deep cosmic void where the density contrast $\delta$ is already approaching its lower bound $-1$) the evolutionary contribution $\Delta {\cal T}_{\cal E}$ can have the same sign as the time-delay contribution $\Delta {\cal T}_{\cal T}$ and consequently considering only the $\Delta {\cal T}_T$ contribution might under-estimate the total signal.
Recent observations show some evidence \cite{Sutter12, Lavaux12, Hamaus14} for the existence of cosmic voids much deeper than predicted by linear theory.
There are theories of structure formation that  predict such deep voids \cite{Ostriker81, Fillmore84, Bertschinger85}.
It is important to study the ISW/RS effect in such strongly nonlinear growth periods to obtain an accurate understanding of what nonlinear effects are possible.
In this paper, we present two new void lens models which include nonlinear growth of the lensing structures, both of which can produce photometric hot rings in the outer regions of the lens.
\section{Two Evolving Embedded Void Lens Models}

Any embedded lens can be thought of as a rearrangement of the mass $M$ in a comoving sphere of the homogeneous background cosmology.
We have constructed probably the simplest embedded lens models in \cite{Chen13a} to study the ISW effect caused by cosmic voids and galaxy clusters (central under and over-densities, respectively).
The simple void lens in \cite{Chen13a} contains two components: an under-dense homogeneous dust sphere of density $\rho = (1-\xi)\bar{\rho}$\  containing a total mass $(1-\xi)M$ surrounded by a compensating over-dense thin shell containing the compensating mass $\xi M$.
Here  $\bar{\rho}$ is the cosmic mean density at the lens redshift and the parameter $\xi=-\delta>0$ characterizes the void's depth: the closer  $\xi$ is to 1, the deeper it is.
Similarly simple void density profiles have been used previously to model void evolution and to study weak lensing by cosmic voids \cite{Maeda83a,Maeda83b,Amendola99}.
For example, the inverted top-hat void model compensated by an over-dense thin shell of finite thickness has been used in \cite{Amendola99} to model cosmic voids as gravitational lenses.
Very simple void profiles have also been used by other authors to study the ISW effect \cite{Rudnick07, Das09}.
To accurately compute the ISW effect produced by a cosmic void, linear lensing theory requires that we know the void's density profile and its first time derivative (but no higher) at the time it lenses the CMB.
When estimating the ISW effect caused by voids of such simple density profiles in \cite{Chen13a}, we assumed two evolving scenarios: one where the lens was co-expanding with the background FLRW universe for which only the lensing time delay contributes to the CMB temperature perturbation, and a second where the density contrast $\delta$ was evolving according to linear structure formation theory.
The simple embedded void  model developed in \cite{Chen13a} predicts $\Delta {\cal T}\le0$ across the whole void.
For that model there is no hot ring $\Delta {\cal T}>0$ surrounding the cold spot as observed by \emph{Planck} or predicted by numerical simulations \cite{Cai10}.
In this Section we extend the model of \cite{Chen13a} to include an evolutionary part, and estimate evolution's possible effect on the ISW signal.
It is important to recall that we are measuring evolution relative to the expanding background Universe.

\subsection{Embedded Lens Model I---A Snowplow}

As with all embedded lenses the outer physical radius $r_d$ of the lens is related to its comoving radius $\chi_b$ by $r_d= R(t_d)\chi_b$  at the time $t_d$ (the time the photons encountered the deflecting lens on their way to be observed at $t_0$) and where $R(t_d)$ is the radius of the homogeneous universe at $t_d$.
Because the mass density of the background cosmology
 $\bar{\rho}(t_d)\propto R(t_d)^{-3}$, the net mass $M$ contained within the sphere is constant (to lowest order in the curvature) and equals $ (4/3)\pi r_d^3\, \bar{\rho}(t_d)$.
For this particular embedded void model the mass  within the sphere is rearranged into two concentric components: a homogeneous under-dense interior of density $(1-\xi)\bar{\rho}$ and radius $r_{\rm v}<r_d,$ and an over-dense infinitesimally thin shell at $r_{\rm v}$ which completely compensates the under-dense interior.
The remainder of lens mass in the outer shell $r_{\rm v}<r<r_d$ remains the same as the background cosmology in which it is embedded, i.e., is of density $\bar{\rho}(t_d)$.
We assume the thin shell of radius $r_{\rm v}$ is propagating (like a shock front) outward.
Consequently the (normalized) radius of the shell $y(t_d)\equiv r_{\rm v}/r_d<1$ increases with time, and the thin shell gains  mass as it plows into $\bar{\rho}$. We can also have the shell contract and lose mass as it fills the lower density inner void region, even though such motion seems rather un-physical.
An over-dense expanding thin shell that surrounds an under-dense interior was proposed by several void formation scenarios, e.g., via
explosive blast waves \citep{Ostriker81,Fillmore84,Bertschinger85}.
The infinitesimally thin shell approximation has also been used to model void formation or lensing by cosmic voids \citep{Maeda83a,Maeda83b,Chen13a,Chen13b}.
The thin shell approximation is made for simplicity and replacing it by a shell of finite thickness is straightforward \cite{Kantowski14}.
In this model any light ray that passes outside the compensating shell at $y(t_d)$ is un-affected by the lens.
As a consequence the actual size of the outer boundary of the comoving sphere $r_d$ is unimportant as long as $r_{\rm v}$ does not overtake $r_d$ as the CMB photons pass through the lens.
We now investigate how the ISW signals produced by this model where  $y(t_d)=r_{\rm v}/r_d$ is allowed to vary, differ from those of the co-expanding lens model presented in \cite{Chen13a}
where $y(t_d)$ was kept constant.
In what follows it is necessary to think of evolving lens parameters, such as the expanding shell's fractional radius $y$, sometimes as functions of cosmic time $t_d$ at lensing and sometimes as functions of the lens's redshift $z_d$, i.e., sometimes as $y(t_d)$ and sometimes as $y(z_d)$. The two independent parameters are related by $1+z_d=R_0/R(t_d)$ and obviously a quantity that increases with $t_d$ appropriately decreases with $z_d$ and vice-versa.

It is straightforward to compute the projected mass profile of this spherical void lens
\be\label{fx_model1}
f(x,z_d)-f_{\rm RW}(x) = \begin{cases}
   &  -\xi x^2 [y^2(z_d)-x^2]^{1/2},  \text{  $0\le x\le y,$} \\
   & 0,    \text{\hspace{82pt}  $y< x \le 1$}.
\end{cases}
\ee
This lens profile is evolving with respect to the FLRW background because of the time (or equivalently the redshift) dependence of the shell's radius $y.$
The potential part of the gravitational lensing time delay is obtained using Eqs.~(\ref{Tp}) and (\ref{fx_model1})
\bea
cT_p 
&=& 2(1+z_d)r_{\rm s}\left[-\frac{\xi}{3}(y^2-x^2)^{3/2}\right], \hspace{10pt} 0\le x\le y,
\eea
and the CMB temperature perturbation cross such a lens is [see Eq.~(\ref{dT})]
\be\label{dT_model1}
\frac{\Delta {\cal T}}{\cal T} 
 =-\frac{2\xi}{3}\frac{r_{\rm s}}{c/H_d}(y^2-x^2)^{1/2}\Bigg[(y^2-x^2) - \frac{3yv}{r_dH_d} \Bigg], \> 0\le x\le y,
\ee
where the redshift derivative needed in Eq.\,(\ref{dT}) was related to the time derivative by
\be
\frac{d\ }{dz_d}=-\frac{1}{(1+z_d)H_d}\ \frac{d\ }{dt_d},
\ee
and where we have defined $v\equiv r_d\ \dot{y}(t_d)$ to be the physical velocity of the shell with respect to the FLRW background, i.e., the radial peculiar velocity.
Both $T_p$ and $\Delta {\cal T}$ vanish outside the over-dense thin shell at $y(t_d).$
In Eq.\,(\ref{dT_model1}) the two terms within the square brackets correspond to the time-delay and evolutionary contributions, respectively.
They can be of the same or opposite sign depending on whether the shell is contracting or expanding.
As expected this result is independent of the choice of $r_d$  since $r_{\rm s}\propto r_d^3$ and both $x$ and $y$ are $\propto r_d^{-1}$.
Without any loss of generality,  we take $y=1$ (after $z_d$ differentiation) from now on which simplifies Eq.\,(\ref{dT_model1}) to
\be\label{T_shock}
\frac{\Delta {\cal T}}{\cal T} =-\frac{2\xi}{3}\frac{r_{\rm s}}{c/H_d}(1-x^2)^{1/2}\Bigg[(1-x^2) -  \frac{3v}{r_dH_d} \Bigg],
\ee
and which reduces to Eq.\,(13) of \cite{Chen13a} when $v=0.$
With this choice, $r_{\rm s}$ is the Schwarzschild radius of the total mass contained within the sphere of radius $r_{\rm v}(t_d)$, including the compensating shell at the time the photons transit the lens.
Equation\,(\ref{T_shock}) for $\Delta {\cal T}$ now contains a term explicitly dependent on the velocity of the propagating shock front \cite{Birkinshaw83,Cooray02,Schafer06,Merkel13}.
In Section~\ref{sec:examples} we will use this equation to model the ISW cold spot across evolving cosmic voids possessing radially diverging (or possibly converging) flows.

\subsection{Embedded Lens Model II---Void with A Running Wall}

For model II we assume the lens has a physical radius $r_d=R(t_d)\chi_b$ and contains  three mass components.
The first is a small point mass
$\eta M\ll M$ at the center of the void (a large $\eta$ value would produce an embedded cluster model, see \cite{Chen13a}).
This small (with respect to the total mass $M$ of the lens) point mass was introduced to represent substructures
near the center of a cosmic void (e.g.,  a galaxy group or a compact object such as a black hole), and to possibly explain the strange small positive excesses in temperature appearing at the centers of stacked photometric profiles \cite{Granett08,Ilic13,Planck14a} when filter sizes below $\sim$ 0.2 times the voids' effective radii are used.
Including a small central mass
was also motivated by the fact that void-finding algorithms such as ZOBOV use galaxies as centers of the tessellation, and consequently the stacked void centers can be slightly over-dense \cite{Neyrinck08,Sutter12}.
The second component of this void model is a moving thin shell
at fractional radius  $y(t_d)<1$ which contains a constant total mass $\xi M \le (1-\eta )M.$
The third component is a uniform density $(1-\eta-\xi)\bar{\rho}\ge 0 $ for all $x<1$,
i.e., of constant density contrast $\delta=-(\eta+\xi)$.
The thin shell can be moving out or in at the time the CMB photons pass, i.e., can be expanding or contracting with respect to the background.
The mass of the shell is assumed not to change whether expanding or contracting, in contrast to model I.
Without the central point mass the lens should behave somewhat like a cluster lens (over-density) outside of the shell and possibly produce a hot ring there and behave like a void lens (under-density) near the center of the lens and possibly produce a central cold spot there.

The projected mass profile of this lens is
\bea
f(x,z_d)-f_{\rm RW}(x) &=& (\eta+\xi)(1-x^2)^{3/2} \cr
&-&\xi{\Theta(y-x)}\left(1-\frac{x^2}{y^2(z_d)}\right)^{1/2}
\eea where $\Theta(y-x)$ is the Heaviside step function (equals 1 when $y-x>0$ and 0 otherwise).
The potential part of the time delay is
\bea
\text{\hspace{-10pt}}cT_p &=&2(1+z_d)r_{\rm s}\Bigg\{
(\eta+\xi)\Bigg[\ln \frac{1+\sqrt{1-x^2}}{x}-\frac{4-x^2}{3}\sqrt{1-x^2}\Bigg] \cr
&&- \xi\Theta(y-x)\Bigg[\ln \frac{1+\sqrt{1-x^2/y^2}}{x/y} -\sqrt{1-x^2/y^2} \Bigg]\Bigg\},
\eea
which gives an ISW effect across the lens
\bea\label{T_shell}
\frac{\Delta {\cal T}}{\cal T} &=& \frac{H_dT_p}{1+z_d}+\Theta(y-x)\frac{2r_{\rm s}\xi}{c/H_d}\frac{\sqrt{1-x^2/y^2}}{y}\frac{\dot{y}(t_d)}{H_d}\cr
&=&\frac{2r_{\rm s}}{c/H_d}\Bigg\{ (\eta+\xi)\Bigg[\ln \frac{1+\sqrt{1-x^2}}{x}-\frac{4-x^2}{3}\sqrt{1-x^2} \Bigg]\cr
&&-\xi\Theta(y-x)\Bigg[ \ln \frac{1+\sqrt{1-x^2/y^2}}{x/y} -\sqrt{1-x^2/y^2}\cr
&&-\frac{\sqrt{1-x^2/y^2}}{y}\frac{v}{r_dH_d}\Bigg] \Bigg\},
\eea
where $\dot{y}(t_d)\equiv v/r_d$ and the very last term ($\propto  v$) is the evolutionary contribution $\Delta {\cal T}_{\cal E}.$
When $\eta=0,$ $y=1$ and $v=0,$ Eq.\,(\ref{T_shell}) reduces to Eq.\,(13) of \cite{Chen13a}.

\section{Examples and Discussions}\label{sec:examples}

We now illustrate the size and shape of the ISW signals produced by the two lens models given in the previous Section [i.e., Eqs.\,(\ref{T_shock}) and (\ref{T_shell})].
We choose a standard $\Lambda$CDM cosmology with $\Omega_{\rm m}=0.3,$ $\Omega_\Lambda=0.7$ and Hubble constant $H_0= 70\rm\, km\, s^{-1}\, Mpc^{-1},$ and we choose a lens redshift $z_d=0.5.$
We assume the lens to be either co-expanding with the FLRW background, i.e., $v=0,$ or evolving because the thin shell is expanding or contracting, $v\ne 0$.
The moving shell  gains/loses mass in model I or has a constant mass in model II.
Model I has three parameters: the physical radius $r_{\rm v}=r_d$ of the void, the void deepness parameter $\xi,$ and the velocity $v$ of the moving shell.
Model II has two additional parameters, $y(t_d)$ [with $0\le y(t_d) \le 1$] which is the radius of the moving shell relative to the radius of the lens, and $\eta,$
which is the fraction of the lens mass contained in the central compact object.
For a model I lens we choose $r_d= 50\, \rm Mpc$ with $v =0,$ $\pm500\,\rm km/s,$ or $\pm 1000\,\rm km/s,$ and  $\xi=-\delta = 0.5.$
For a model II lens we choose $r_d=50\,\rm Mpc$, $y(t_d)=0.5$ or 0.8 with $v = 0,$ $\pm 250\, \rm km/s,$ or $\pm 500\, \rm km/s,$ $\xi = 0.9,$ and $\eta=0$ (i.e., no central compact object) or $\eta=0.01$ (a compact central object containing a small fraction of the lens mass).

Given the above assumed numbers we can check the appropriateness of having dropped all post-Newtonian corrections in the gravity perturbations.
The projected perturbed void density profiles we use are smooth and their gravitational effects remain strictly Newtonian even when $\delta\approx -1$.  For example a large void of physical radius $\sim$$50\,\rm Mpc$ at redshift $z=0$ constructed from a $\sim$$2\times10^{16} M_\odot$ Swiss cheese void 
produces a Newtonian perturbation of only $\Phi/c^2 \approx 5\times 10^{-5}$. 
If the shell is moving with a velocity of $\sim$$1000\,\rm km/s$, a post-Newtonian metric correction of a magnitude $\sim$$3\times 10^{-9}$ is generated. This is the same size as $(\Phi/c^2 )^2$, the next higher order curvature correction from GR. Such small terms are never needed.  
A post-Newtonian gravitational correction to the metric can also be generated by a pressure; however, with pressure estimated from the mass motions in our models we find a pressure to energy density ratio $P/(\overline{\rho}\,c^2) \approx 5\times 10^{-5}$. 
Therefore, just as in conventional lensing calculations, pressure effects are negligible.

\begin{figure*}
\begin{center}$
\begin{array}{cc}
\includegraphics[width=0.5\textwidth,height=0.3\textheight]{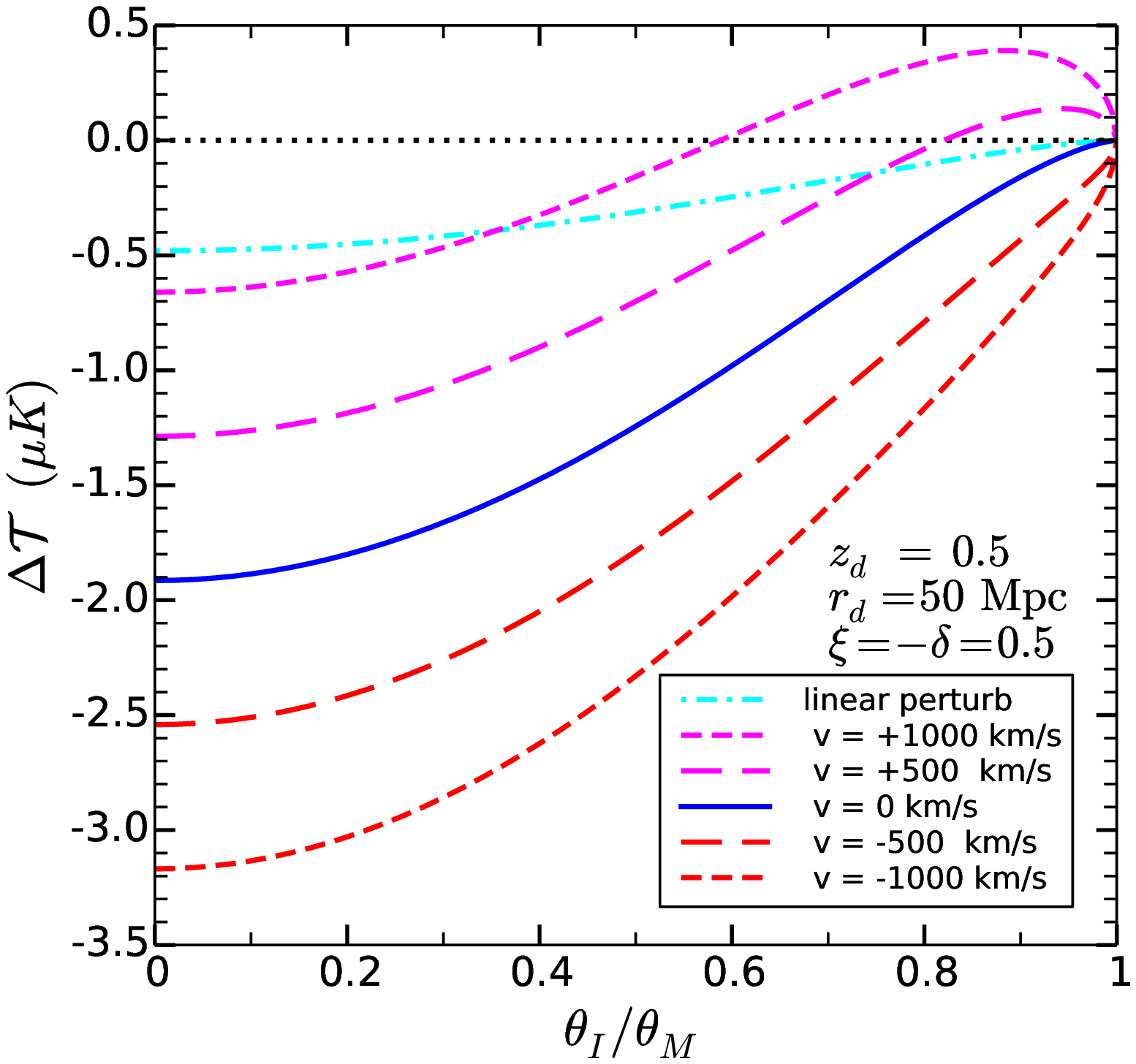}
\includegraphics[width=0.5\textwidth,height=0.3\textheight]{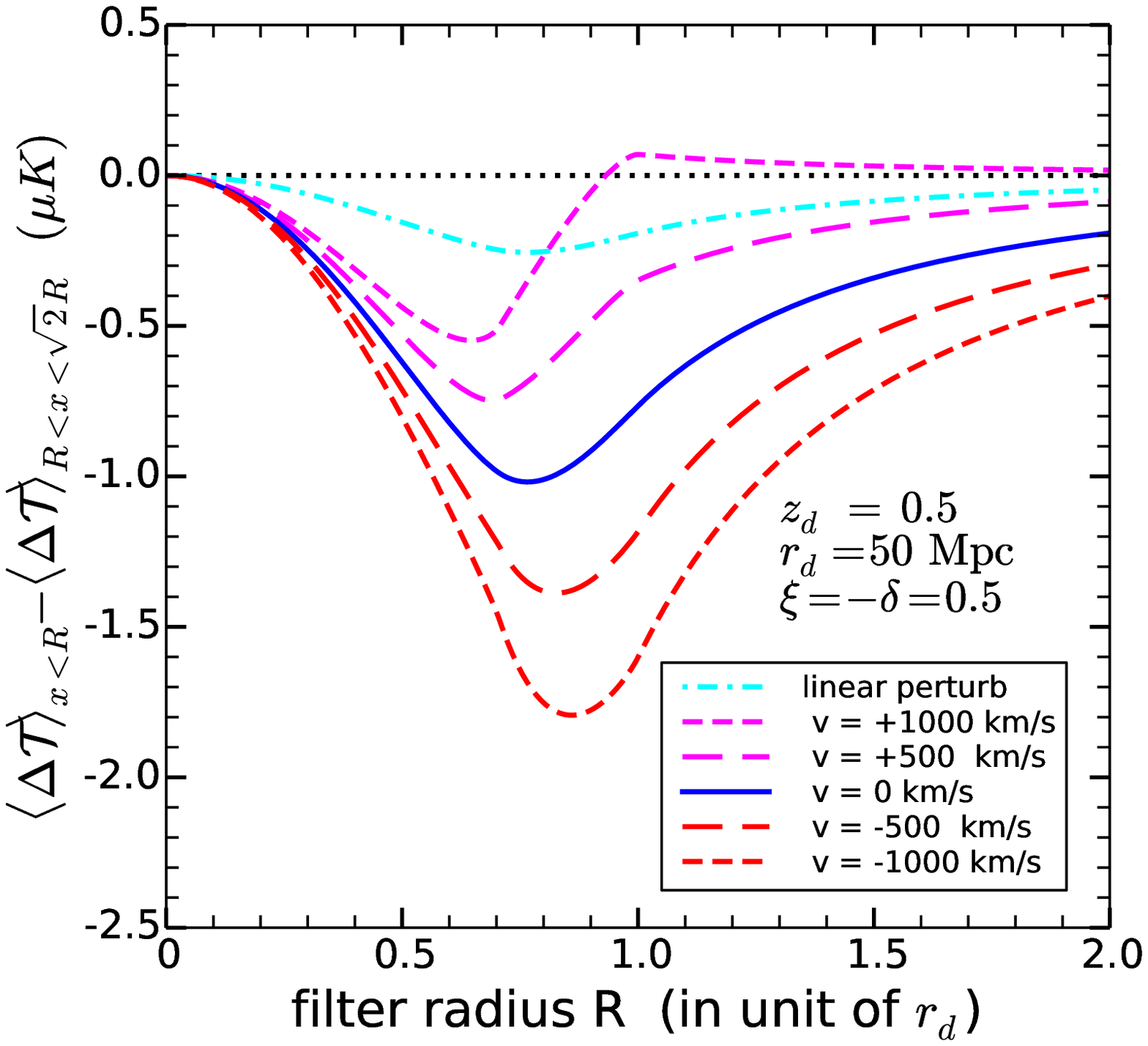}
\end{array}$
\end{center}
\caption{ISW temperature profiles across embedded void lens models of type I are shown in the left panel.
The right panel shows the signals averaged  by applying a compensated top-hat filter, see Eq.\,(\ref{filter}).
The void is at redshift $z_d=0.5$ with physical radius $r_d=50\,\rm Mpc$ and density contrast $-0.5.$
The respective solid, long-dashed, and dashed curves show results for shells which are co-expanding with (blue),  expanding relative to (magenta), and contracting  relative to (red)  the FRLW background.
The relative velocities shown are $\pm1000\,\rm km\,s^{-1}$ and  $\pm500\,\rm km\,s^{-1}.$
The cyan dot-dashed curve is for the case when the shell is co-moving but the density contrast $\delta$ is evolving according to linear growth theory.
A central cold spot is produced for all cases (left panel) but linear growth produces the smallest ISW effect.
Near the void boundary the time-delay contribution diminishes and the ISW signal is dominated by the movement of the compensating shell, which results in  a hot ring for expansion.
On the right the ISW filtered cold spot is seen to be most significant at radius $R\approx 0.6$.
}
\label{fig:model1}
\end{figure*}

\begin{figure*}
\begin{center}$
\begin{array}{cc}
\includegraphics[width=0.5\textwidth,height=0.3\textheight]{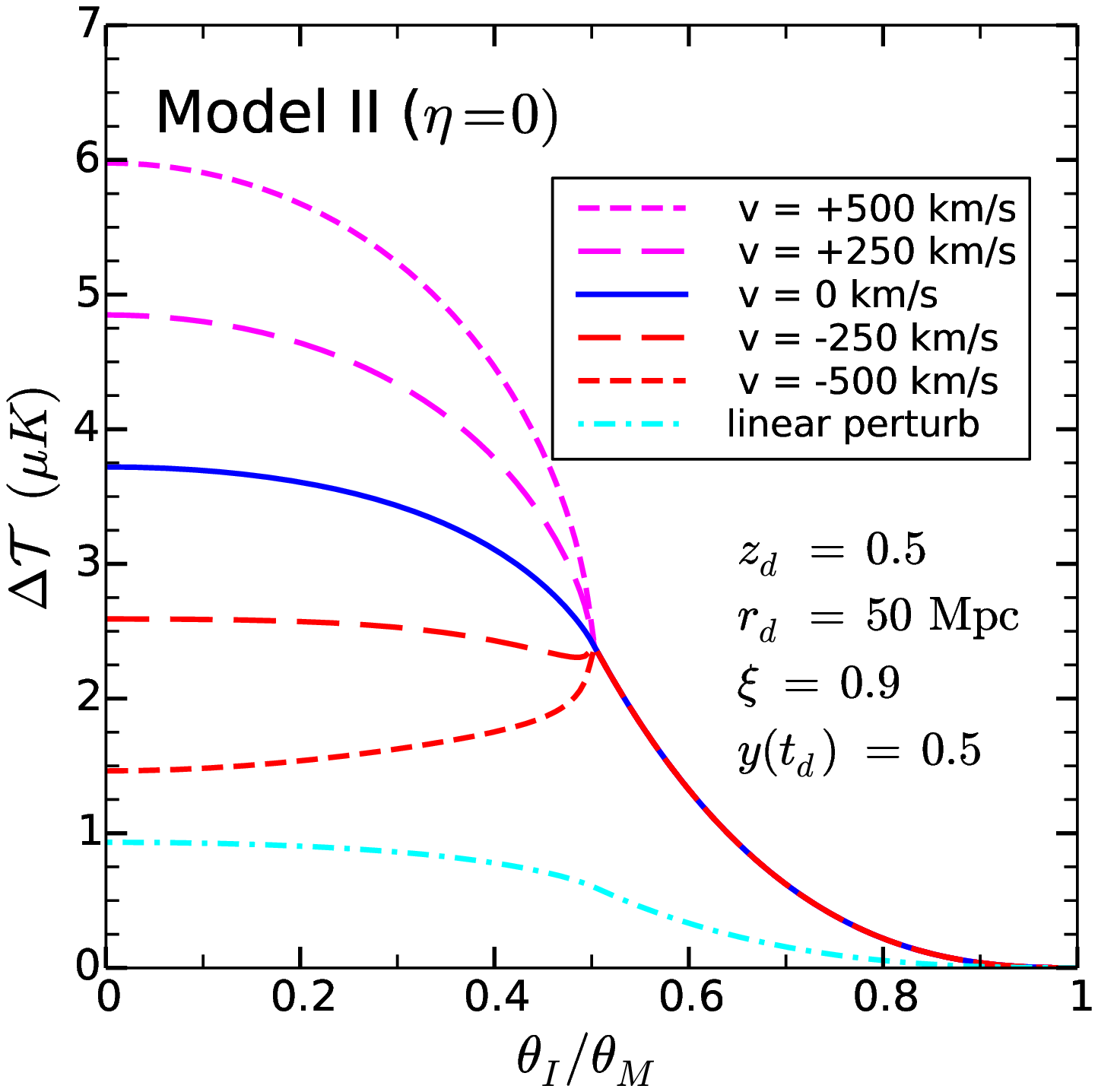}
\includegraphics[width=0.5\textwidth,height=0.3\textheight]{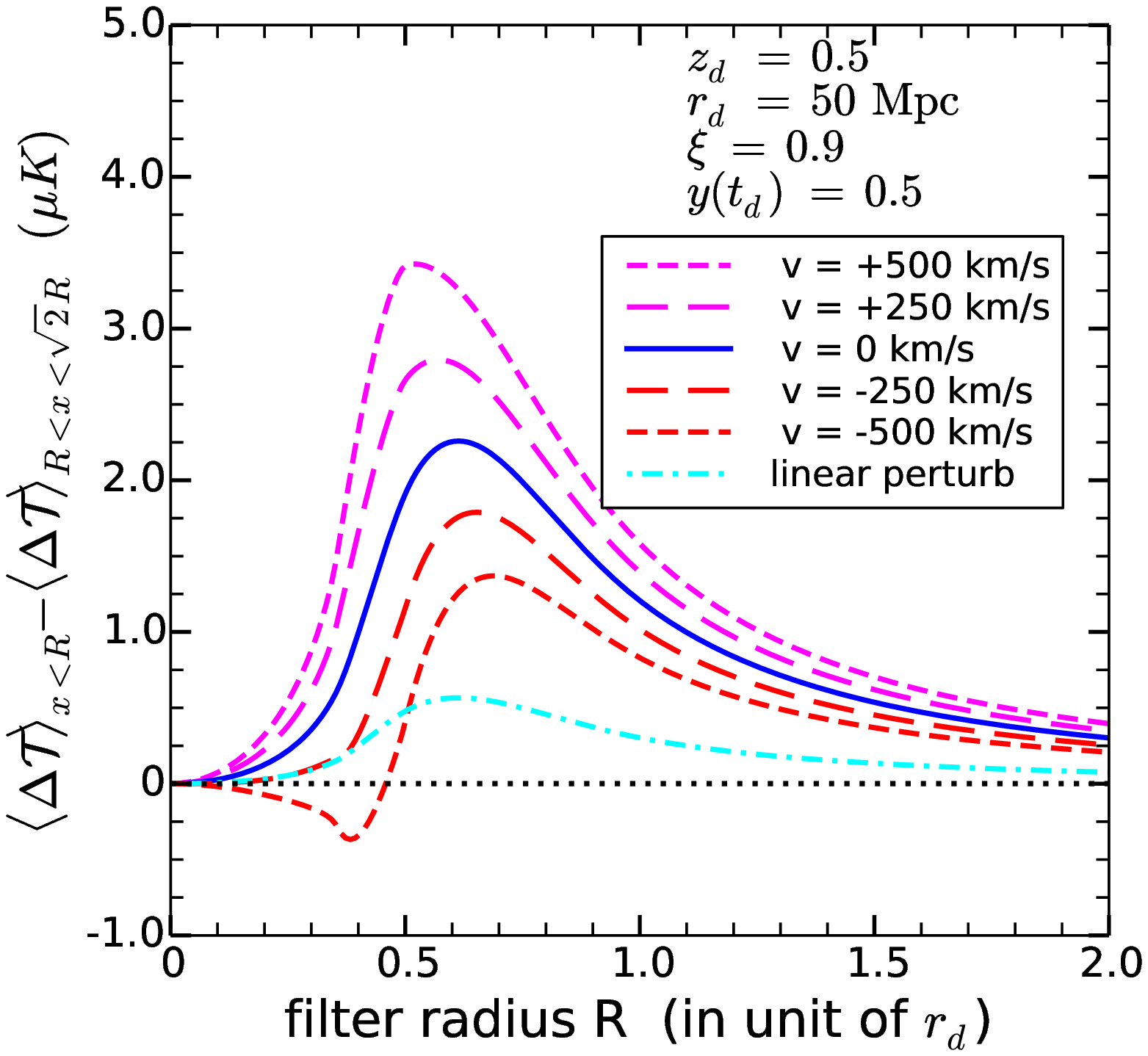}\\
\includegraphics[width=0.5\textwidth,height=0.3\textheight]{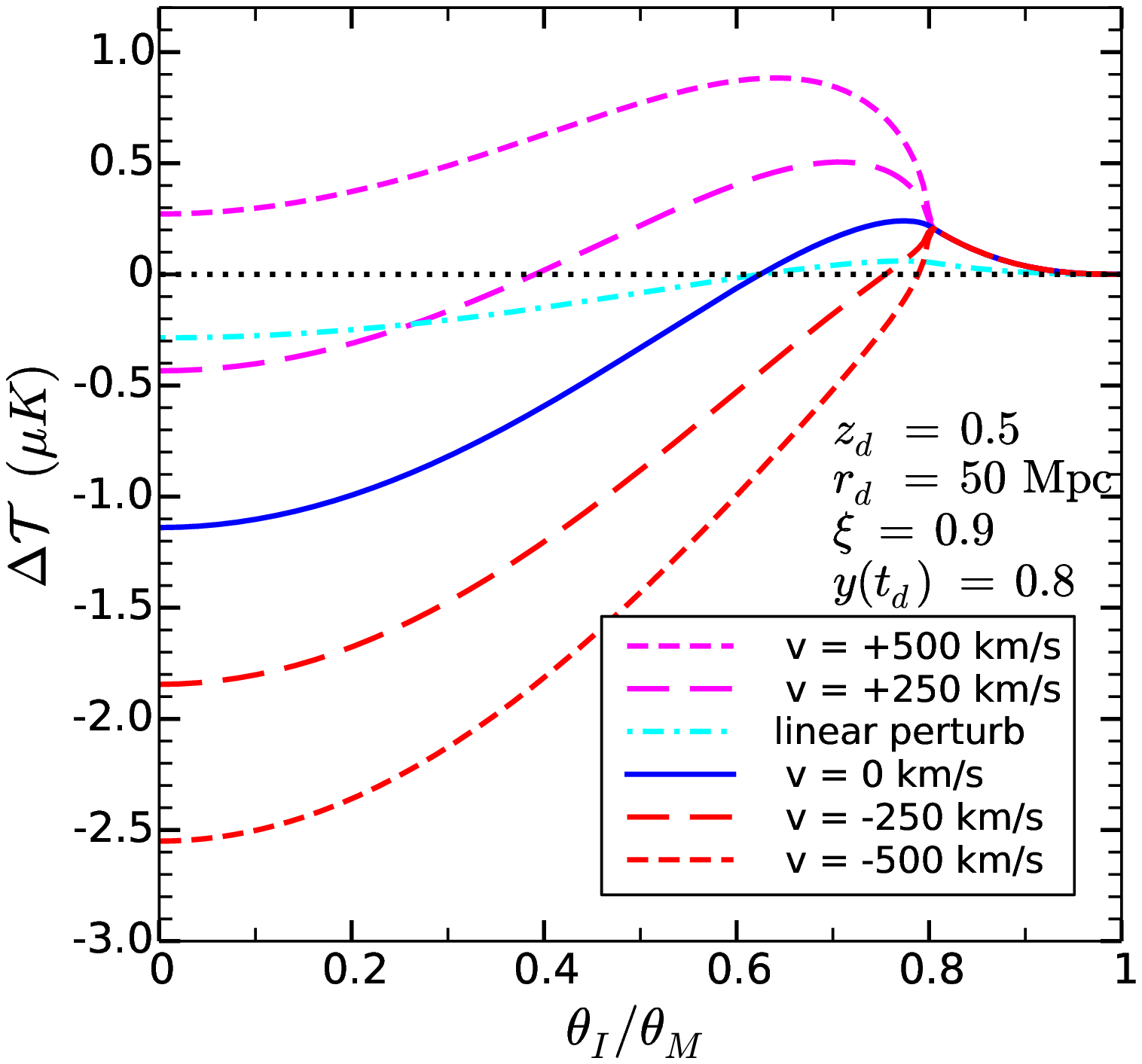}
\includegraphics[width=0.5\textwidth,height=0.3\textheight]{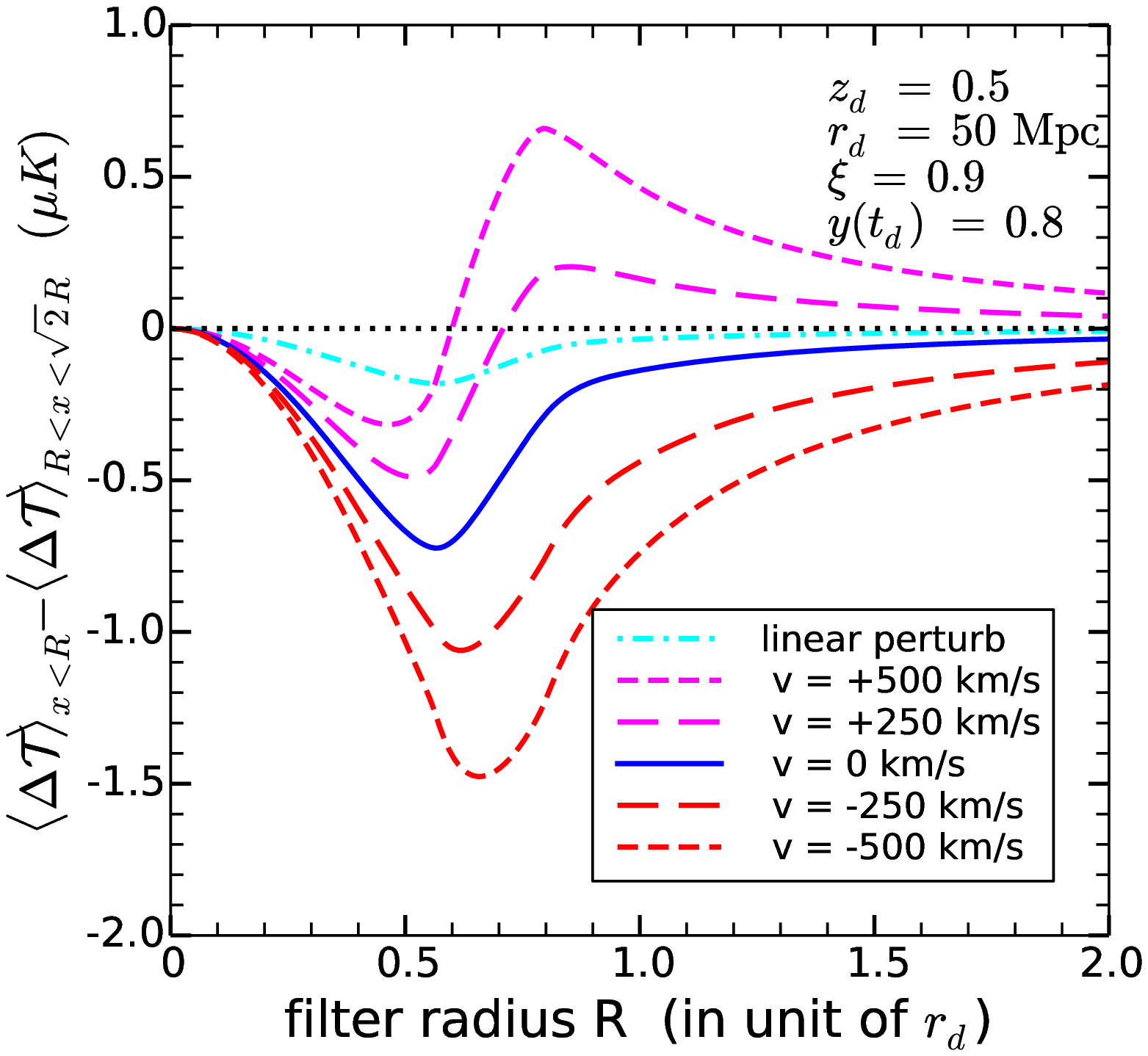}\\
\end{array}$
\end{center}
\caption{  ISW temperature profiles for embedded model II  voids without a central compact object are shown in the left column and the corresponding filtered signals are shown on the right.
The void is at redshift $z_d=0.5$ with physical radius $r_d=50\,\rm Mpc$ and density contrast $\delta = -0.9$.
A compensating shell is located at $y(t_d) = 0.5$ or at  $0.8$ (top or bottom row respectively).
For the upper row the lens behaves like a cluster lens producing a hot spot toward the center.
For the lower row most of the lens mass is located near the void boundary and a central cold spot is produced except for the case of an expanding shell with $v=+500\,\rm km\, s^{-1}$ where $\Delta {\cal T} >0$ across the entire lens.
A diverging/converging flow increases/decreases the CMB temperature but only within the shell (i.e., the solid, dashed, and long-dashed curves merge beyond $x=y(t_d)$).
A hot ring near the void boundary can be produced by this lens model.
The filtered ISW cold spot is again most significant at a filter radius $R \approx0.6$, see the bottom right panel.  }
\label{fig:model2}
\end{figure*}

\begin{figure*}
\begin{center}$
\begin{array}{cc}
\includegraphics[width=0.5\textwidth,height=0.3\textheight]{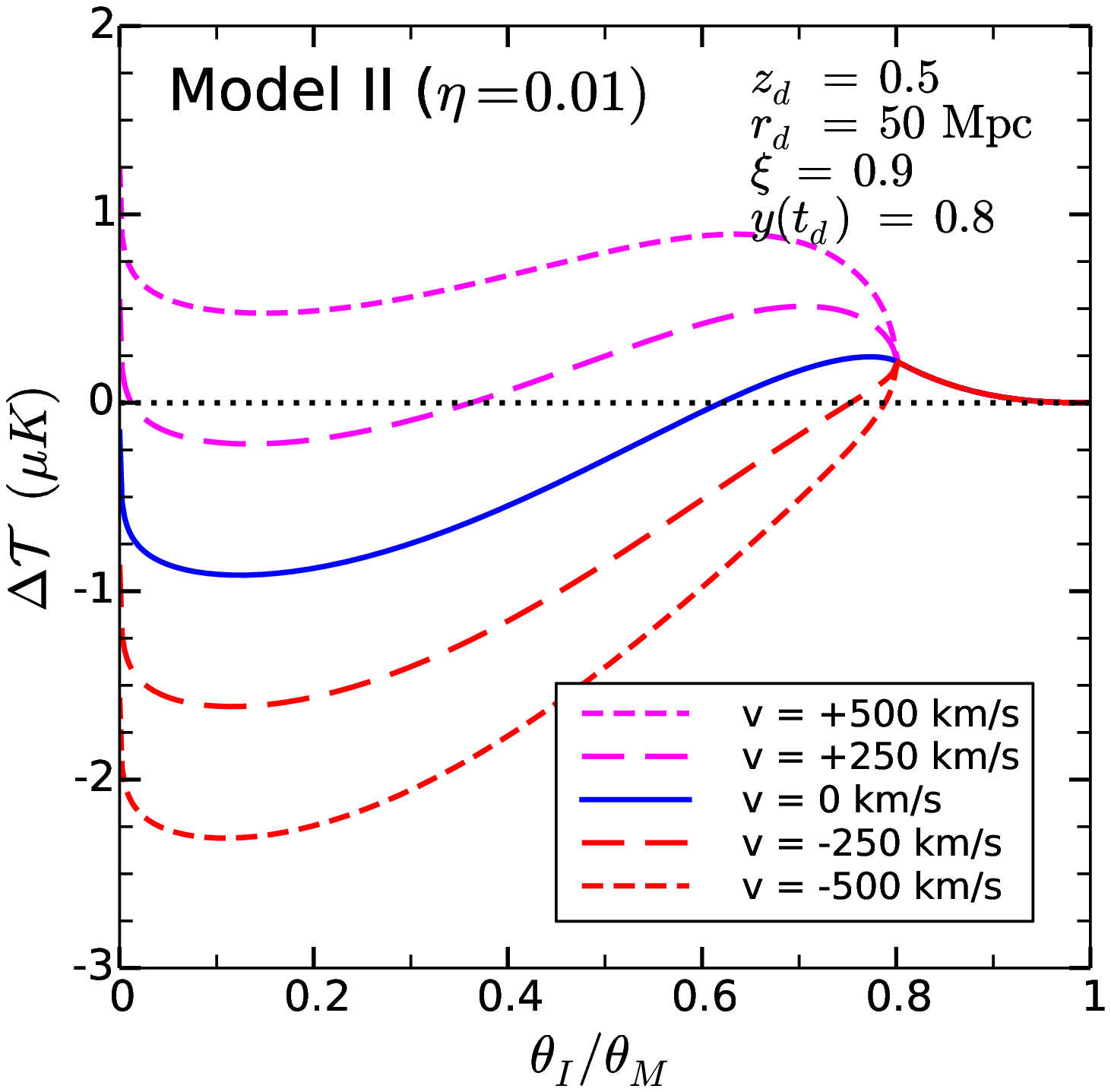}
\includegraphics[width=0.5\textwidth,height=0.3\textheight]{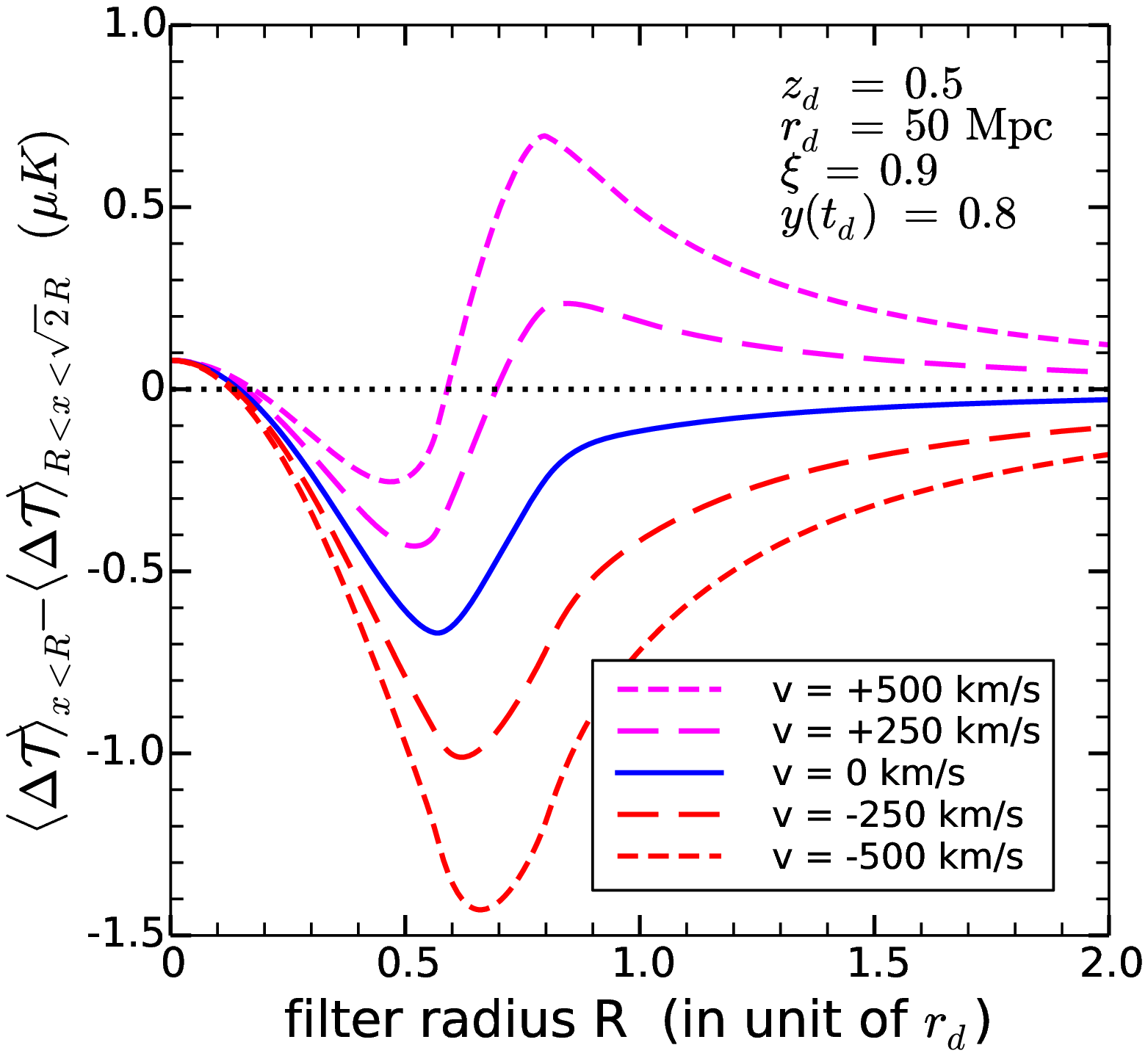}
\end{array}$
\end{center}
\caption{ ISW temperature and photometry profiles (left and right panels respectively) for embedded model II voids possessing a central compact object.
The void is at redshift $z_d=0.5$ with physical radius $r_d=50\,\rm Mpc.$
The over-dense shell is located at $y(t_d)=0.8$ and contains 90\% of the lens mass ($\xi=-0.9$).
The central compact object contains 1\% of the lens mass (about $7.2\times 10^{14} M_\odot$).
Hot rings again appear near the boundary of these large void models and the central cold spot is again most significant at filter radius $R\sim0.6.$
However, now a  small central positive excess is  present for filter radii $R\lesssim0.2$ due to the point mass at the void's center.
}
\label{fig:model2b}
\end{figure*}

The results are shown in Figs.~\ref{fig:model1}--\ref{fig:model2b} respectively for model I, model II without a central object, and model II with a central compact object.
Besides the temperature profile $\Delta{\cal T}(x)$, we also plot the averaged signal as a function of the filter size $R$ using a compensated top-hat filter\footnote{ This compensated filter is often used by observers to reduce the large-scale power contamination from the primordial CMB.}, i.e.,
\be\label{filter}
\langle\Delta {\cal T}\rangle(R) = \frac{2}{R^2}\left[\int_{0}^{R}{T({r})rd{r}}-\int_{R}^{\sqrt{2}R}{T({r})rd{r}}\right].
\ee
The filtered signals are shown in the right panels of Fig.\,\ref{fig:model1} for model I, and  respectively in Figs.\,\ref{fig:model2} and \ref{fig:model2b} for model II without and with a central compact object.

For both lens models, the total ISW signal is proportional to the lens mass ($\propto r_d^3$), and the amplitudes of the signals are of the same order,
$r_{\rm s}H_d/c,$ which is about  $2\times 10^{-6}$ for $r_d=50\, \rm Mpc$ (a lens mass about $7.2\times 10^{16}M_{\odot}$ and $r_{\rm s}\approx 0.007\,\rm Mpc$).
The relative contribution of the evolutionary part with respect to the time-delay part is characterized by $v/(r_dH_d),$ i.e., the ratio of the shell's radial velocity relative to the local Hubble expansion of the void's boundary, and is of order $\sim$ $0.1$ for $r_d= 50\,\rm Mpc$ at redshift $z_d=0.5$ with $v\sim 500\rm\, km/s$ assuming the conventional $\Lambda$CDM cosmology.
For both lens models, the absolute contribution of the evolutionary part is proportional to both the mass and velocity of the moving shell, i.e.,
$\Delta{\cal T}_{\cal E} \propto \xi vr_{\rm s}/(cr_d),$ about $2\times 10^{-7}$ for $r_d\approx 50\,\rm Mpc$, $\xi\approx 0.9,$ and $v\sim 500\rm\, km/s.$
Inside the moving shell, $\Delta{\cal T}_{\cal E}$ is positive for expanding shell (diverging flow), and negative for contracting shell (converging flow).
This result appears consistent with the Birkinshaw-Gull effect for CMB temperature fluctuations where a transversely moving concentrated lens mass redshifts CMB photons on the side approached by the moving lens and blueshifts them on the other side \cite{Birkinshaw83,Gurvits86,Cooray02,Schafer06,Merkel13}.
An expanding spherical shell reddens CMB photons passing just exterior to its boundary and blue shifts photons passing just interior. 
The converse is true if the shell is contracting.

For the first model, the signal is dominated by the evolutionary part near the lens boundary where $x$ approaches 1, see Eq.~(\ref{T_shock}).
Consequently, this model always produces a hot ring near the lens boundary when $v>0$.
Furthermore, for expanding shells (the magenta curves, i.e., the top two dashed curves, in the right panel of Fig.\,\ref{fig:model1}) the compensated top-hat filtered ISW cold spot is most significant at a filter radius $R\approx 0.6$ (i.e., $r\approx 0.6 r_d$), consistent with recent observations and simulations \cite{Granett08, Ilic13, Planck14a,Cai14}.
Predictions of the second model depend significantly on both the position and velocity of the shell.
We first discuss the case without a compact object at the void center, i.e., $\eta=0.$
If the shell is located close to the center of the lens (e.g., $y(t_d)\lesssim 0.5$), then the lens behaves like a cluster lens,
see the first row of plots in Fig.~\ref{fig:model2}.
If the shell is located near the boundary of the lens, then the lens behaves like a void lens with a cold spot toward the center,
see the second row of plots in Fig.~\ref{fig:model2}.
However, if the expansion velocity of the shell is high enough (e.g., $v=500\, \rm km/s$), the lens produces a hot spot, $\Delta {\cal T} >0$ across the entire lens, see the dashed magenta curve, i.e., the top curve in the  bottom left panel of Fig.~\ref{fig:model2}.
For model II the motion of the shell only impacts regions inside the shell, i.e., the curves corresponding to an expanding, static, and contracting shell fork within the shell but merge at $y(t_d)$ and beyond.
If $\xi$ is small, then motion of the shell is less important because the shell contains only a small fraction of the lens mass and consequently the lens mass structure, i.e., $f(x),$ will not change significantly even if the shell is moving rapidly.
For comparison, in Fig.\,\ref{fig:model2} we also present the results for the linear growth case (the cyan dotted-dashed curves) where we have assumed the density perturbation $\delta$ to be evolving according to linear structure theory. 
For this case the signal is suppressed  by a factor of about 4 comparing with the static case (refer to Fig.~3 of \cite{Chen13a}).

Figure\,\ref{fig:model2b} shows the results for void model II with a central compact object containing 1\% of the lens mass (about $7.2\times 10^{14} M_\odot$).
In the left panel of Fig.\,\ref{fig:model2b} the spikes at the void's center are caused by the fact that we have assumed a simple point mass for the compact object (refer to \cite{Chen13a} for a more complicated model).
ISW signals averaged  by applying a compensated top-hat filter show a small central positive excess ($\langle\Delta{\cal T}\rangle>0$ for filter size $R\lesssim0.2$), see the right panel of Fig.\,\ref{fig:model2b}.
This is encouraging given that the ISW temperature profile obtained by applying the aperture photometry techniques to the recent \emph{Planck} observations and large void catalogs \cite{Sutter12} did contain a strange (even if with a low significance) positive excess near the center of the stacked images, see Fig.\,9 of \cite{Planck14a}.
This result is consistent with the suggestion made in \citet{Planck14a} that the positive excess may be partially caused by the fact that voids used for the photometry profiles contain small central over-densities.
We find that a void with a moving shell and a central compact object can produce both the hot ring near the boundary of the void and a positive excess near the void's center.

\section{conclusions}



We have recently developed the embedded lens theory and the Fermat potential formalism for studying the ISW effect caused by individual compensated inhomogeneities \cite{Kantowski10, Kantowski12, Kantowski13, Chen10, Chen11, Chen13a, Chen13b, Kantowski14}.
Just as with the conventional lensing potential approach, the Fermat potential approach is based on Newtonian perturbations of the background cosmology and when applied to similar lenses both approaches give similar results. The Fermat potential approach, however, is much simpler because it uses the projected lensing mass directly and bypasses the step of computing the Newtonian lensing potential.
Our method can be used to model the ISW signals extracted by stacking patches of CMB sky maps around known cosmic voids or galaxy clusters \cite{Granett08,Planck14a},
or to model the CMB cold spot on the south hemisphere as an ISW/RS signals produced by large nearby cosmic voids \cite{Rudnick07, Velva04, Das09, Finelli14, Szapudi14}. The current modeling difficulty is the uncertainty in the mass density profile of voids and its dynamics.
The stacked radial void profiles obtained from large void catalogs \cite{Sutter12,Pan12} built from galaxy redshift surveys such as the  Sloan Digital Sky Survey \cite{Abazajian09} show deep interiors toward the voids' centers ($\delta\lesssim-0.9$) for voids of radii from a few up to about $100\,\rm Mpc.$ These profiles also show compensating over-dense bounding ridges (see e.g., Figure 9 of  \cite{Sutter12}).
The thickness and profile of the over-dense bounding regions, as well as the extent to which they compensate the under-dense regions are not currently very well constrained by observations.
An additional problem with modeling mass densities occurs  because luminous matter as a tracer of dark matter is biased, i.e., the underlying void dark matter profile may be much shallower than estimates based on galaxy counts.
Another modelling complication arises if
suggestions that large voids tend to be under-compensated whereas small voids might be over-compensated \cite{Ceccarelli13,Hamaus14} are true.
The universal void dark matter density profile of \cite{Hamaus14} based on $\Lambda$CDM N-body simulation suggests a deep void interior, $\delta \approx -0.95$ for small voids of radius $\sim$$10\rm h^{-1}$ Mpc, and $\delta \approx -0.5$ for large voids of radii $\sim$$70\rm h^{-1}$ Mpc.
These dark matter profiles are much deeper than those predicted by linear growth theory so that estimating densities based on the linear theory will be in error.
Rather than attempting to model voids with complicated uncertain analytic expressions or with numerically motivated density profiles we have applied our Fermat based calculation of the ISW  effect, Eq.\,(\ref{dT}), to two simple embedded lens models for the purpose of  illustrating the effect evolving densities can have on a void's temperature profile.
These models possess deep voided regions and nonlinearly developing mass shells that produce hot rings around central cold spots
and can thus explain recent observations found using the aperture photometry technique \citep{Granett08,Planck14a}.
The simplicity of the profiles (e.g., a uniform under-dense interior with an infinitesimally thin over-dense shell moving outward or inward) combined with Eq.\,(\ref{dT}) allows us to give results in analytical form, and shed light on how converging/diverging flows influence the CMB observations.
We have assumed these void lenses to be strictly compensated with net mass zero with respect to the FLRW background.
When presenting examples of the two simple void lens models, we have chosen voids of radii $50\,\rm Mpc$, a size not unusual according to recent void catalogs \cite{Sutter12,Pan12,Nadathur14}.
The 50 large voids used in \cite{Granett08} have a mean redshift of about 0.5 and an average radius of about $100\,\rm Mpc.$
The assumed void density contrast, $\delta=-0.5$ or $-0.9$ (model I and II, respectively) was largely based on the stacked void density profiles of \citep{Sutter12}.
While void redshifts and radii are provided by void catalogs, good estimates for compensating shell thicknesses and their expansion velocities
are lacking.
For simplicity, we have assumed expanding thin shells with velocities of the order of a few hundred kilometers per second, i.e., a magnitude similar to that of the peculiar velocities of galaxies.

Expanding the analytical work given here to more realistic profiles and/or more complicated flows \citep{Lavaux12,Sutter12,Hamaus14,Cai10,Cai14} is straightforward but awaits observational constraints, and may have to be done numerically.
We conclude that the embedded lens theory and the formalism of Fermat potential is well suited for modeling the ISW effect and that mass motions within voids  can easily produce hot and cold rings. Mass profiles with a central compact object can even produce a central positive temperature excess as observed in \cite{Planck14a}.

\begin{acknowledgments}
The authors would like to thank the anonymous referee for a careful review of this work.
\end{acknowledgments}

\label{lastpage}

\end{document}